\documentclass[twoside,leqno,twocolumn]{article}

\usepackage[letterpaper]{geometry}

\usepackage{ltexpprt}

\def\PP{{{\rm l}\kern - .15em {\rm P} }}
\def\PN2{{\PP_{N}-\PP_{N-2}}}

\newcommand{\bphi}{\boldsymbol{\varphi}}

\newcommand{\btau}{\boldsymbol{\tau}}

\newcommand{\ba}{\boldsymbol{a}}

\newcommand{\bu}{\boldsymbol{u}}

\newcommand{\deleted}[1]{{}}

\usepackage{algorithm}
\usepackage{algorithmic}
\usepackage{amsfonts,psfrag,amsmath,bbm,color}
\usepackage{graphicx}
\graphicspath{{./figure/}}
\usepackage{amscd}
\usepackage{url}

\begin{document}

\title{\Large Closure Learning for Nonlinear Model Reduction Using Deep Residual Neural Network}
\author{Xuping Xie\thanks{xiex@ornl.gov, Oak Ridge National Laboratory.}
\and Clayton Webster\thanks{webstercg@ornl.gov, Oak Ridge National Laboratory.}
\and Traian Iliescu\thanks{iliescu@vt.edu, Virginia Tech.}}

\date{}

\maketitle





\fancyfoot[R]{\scriptsize{Copyright \textcopyright\ 2020\\
Copyright retained by principal author's organization}}

\pagenumbering{arabic}
\setcounter{page}{1}

\begin{abstract} \small\baselineskip=9pt Developing accurate, efficient, and robust closure models is essential in the construction of reduced order models (ROMs) for realistic nonlinear systems, which generally require drastic ROM mode truncations.
We propose a deep residual neural network (ResNet) closure learning framework for ROMs of nonlinear systems.
The novel ResNet-ROM framework consists of two steps:
(i) In the first step, we use ROM projection to filter the given nonlinear PDE and construct a filtered ROM.
This filtered ROM is low-dimensional, but is not closed (because of the PDE nonlinearity).
(ii) In the second step, we use ResNet to close the filtered ROM, i.e., to model the interaction between the resolved and unresolved ROM modes.
We emphasize that in the new ResNet-ROM framework, data is used only to complement classical physical modeling (i.e., only in the closure modeling component), not to completely replace it.
We also note that the new ResNet-ROM is built on general ideas of spatial filtering and deep learning and is  independent of (restrictive) phenomenological arguments, e.g., of eddy viscosity type.
The numerical experiments for the 1D Burgers equation show that the ResNet-ROM is significantly more accurate than the standard projection ROM.
The new ResNet-ROM is also more accurate and significantly more efficient than other modern ROM closure models.
\end{abstract}
\section{Introduction}
Many scientific and engineering applications, such as weather forecasting, ocean modeling, and cardiovascuar flow simulation, often can be represented by a high-dimensional ordinary differential equation (ODE) or partial differential equation (PDE) system.  Analysis and high-fidelity simulation of such a system are very expensive even with modern supercomputer using thousands of cores. Consequently, the use of full system for such simulations is impractical and prohibitive for time critical applications, such as flow control and parameter estimation, where repeated full order model (FOM) simulations are required. To alleviate the computation burden of the FOM simulation, reduced order models (ROMs) have been successfully used.  

ROMs seek a low-dimensional approximation of the FOM with orders of magnitude reduction in computational cost. The classical projection based ROM approach is first to construct a low dimensional space using data-driven reduction methods, such as proper orthogonal decomposition (POD) or dynamical modal decomposition (DMD). The ROM dynamics can be obtained via Galerkin projection of the FOM onto the reduced space. 

The Galerkin projection reduced order model (GP-ROM) is very efficient and relatively accurate for linear systems. However, for realistic nonlinear systems (e.g., convection dominated fluid flows), the GP-ROM can generate inaccurate approximations. 
The main reason is that, due to the inherently drastic mode truncation required in realistic settings,  the dimension of the GP-ROM space is too low to resolve the complex nonlinear interactions of the fluid system~\cite{loiseau2018constrained,noack2016recursive}. 
Current GP-ROMs generally fail to represent the interaction between the GP-ROM modes and the discarded modes, i.e., they do not include a ROM closure model.
Thus, in realistic nonlinear systems, current GP-ROMs not endowed with a closure model yield inaccurate results, often in the form of spurious numerical oscillations~\cite{amsallem2012stabilization,carlberg2013gnat,noack2011reduced}.
This drawback limits the applicability of GP-ROM in many fluid mechanics applications, such as flow control, climate modeling, and weather forecasting~\cite{wang2011two,bergmann2009enablers}

\subsection{Related Work}

ROM closure models for nonlinear systems have been proposed in, e.g.,~\cite{baiges2015reduced,chekroun2017markovian,feppon2018dynamically,hijazi2019data,lu2017data,majda2012physics,san2018neural,wang2012proper,wells2017evolve}. 
The vast majority of the current ROM closure models aim at mitigating the numerical instability observed in GP-ROMs that do not include a closure model.
Some of these ROM closure models use stabilization techniques that have been developed in standard discretization methods (e.g., in the finite element community)~\cite{bergmann2009enablers,carlberg2011efficient,parish2019adjoint}. 
Other ROM closure models have imported ideas developed in standard CFD methodologies, e.g., large eddy simulation (LES)~\cite{HLB96,wang2012proper}.
The overwhelming majority of the current ROM closure models can be categorized as stabilization techniques  (for a notable exception, see the approximate deconvolution ROM closure model~\cite{xie2017approximate} that uses a mathematical framework inspired from image processing).

This is in stark contrast with LES, where a wide variety of closure models have been proposed over the years.
The main difference between ROM closure and LES closure is that the latter has been entirely built around physical insight from the statistical theory of turbulence (e.g., energy cascade and Kolmogorov's theory), which is generally posed in the Fourier setting~\cite{Pop00,Sag06}. 
This physical insight is not available in the ROM setting (see, however,~\cite{CSB03} for initial steps). 
Thus, current ROM closure models have been deprived of this powerful methodology that represents the core of most LES closure models.


\subsection{Our Approach}

Our vision is that machine learning represents the perfect solution for ROM closure modeling.
Indeed, since physical insight cannot be used in a ROM setting, available data and machine learning can be utilized  instead to develop ROM closure models.

\begin{figure}[h]
\includegraphics[scale=0.5]{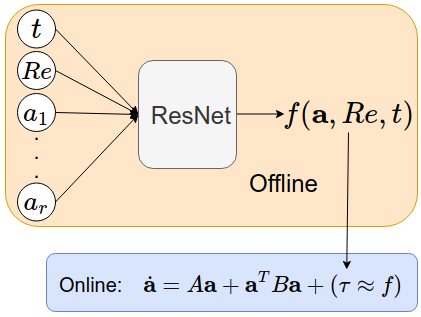}
\caption{
	Flow chart of our model.
	\label{fig:flow-chart}
	}
\end{figure}

We propose a novel ROM closure modeling framework that is constructed by using available data and deep residual neural network (ResNet). 
The resulting ROM, which we call the {\it residual neural network ROM (ResNet-ROM)}, is schematically illustrated in~\eqref{eqn:cl-rom-schematic} and Fig.~\ref{fig:flow-chart} (see Section~\ref{sec:closure}
 for details).
We emphasize that, in the new ResNet-ROM framework, data is used only to complement classical physical modeling (i.e., only in the closure modeling component)~\cite{maulik2019time,xie2018data}, not to completely replace it~\cite{rahman2019non}.
Thus, the resulting ResNet-ROM combines the strengths of both physical and data-driven modeling.
\begin{equation}
\boxed{
\begin{CD}
	\text{FOM}		
	@>
	\text{filtering} 
	>>	
	\text{G-ROM}
	@>
	\text{learning closure} 
	>>	
	\text{ROM} \\[0.1cm]
\end{CD}
}
	\label{eqn:cl-rom-schematic}
\end{equation}
The main contributions of this paper can be summarized as follows:
\begin{itemize}
    \item A novel ROM closure learning framework centered around deep neural networks. 
    \item A hybrid framework that synthesizes the strengths of physical modeling and data-driven modeling.
    \item Very good performance in numerical tests, both in the reconstructive and the predictive regime.
    \item Dramatic improvement in numerical accuracy compared with state of the art ROM closure models.
\end{itemize}

\section{Reduced Order Model}

To construct the novel ResNet-ROM framework, we use the classical Navier-Stokes equations (NSE) that model  incompressible fluid flows:
\begin{equation} 
  \label{eqn:nse}
      {\bu}_{t} - Re^{-1} \, \Delta \bu +(\bu \cdot \nabla)\, \bu + \nabla p = 0 ,
\end{equation}
which are defined on a spatial domain $\Omega$ and on the time interval $[0,T]$.
In~\eqref{eqn:nse}, $\bu$ is the velocity, $p$ the pressure, and $Re$ the Reynolds number. 
The ROM basis $\{ \bphi_1, \ldots, \bphi_r \}$, where $r$ is small, represents the large, energy containing structures in the flow, and is obtained by using available numerical or experimental data and, e.g., the POD or DMD methods. 
The ROM velocity approximation is defined as
\begin{equation}
	{\bu}_r({\bf x},t) 
	\equiv \sum_{j=1}^r a_j(t) \bphi_j({\bf x}) \, ,
	\label{eqn:g-rom-1}
\end{equation}
where $\{a_{j}(t)\}_{j=1}^{r}$ are the sought time-varying coefficients, which are determined by solving the following system of equations: $\forall \, i = 1, \ldots, r,$
    \begin{eqnarray*}
        \left(
            \frac{\partial \bu_r}{\partial t} , \bphi_{i}
        \right)
        + \frac{1}{Re} \, \left(
            \nabla \bu_r ,
            \nabla \bphi_{i}
        \right)
        + \biggl(
            (\bu_r \cdot \nabla) \, \bu_r ,
            \bphi_{i}
        \biggr)
        = 0 .
    \label{eqn:g-rom-weak}
    \end{eqnarray*}
To derive the above equation, we assumed that the ROM velocity modes are perpendicular to the discrete pressure space, 
which is the case if standard mixed FEs (e.g., Taylor-Hood) are used for the snapshot creation. 
Using~\eqref{eqn:g-rom-1}, yields the standard {\it Galerkin projection ROM (GP-ROM)}:  
\begin{eqnarray}
	\dot{\ba}
	= A \, \ba
	+ \ba^{\top} \, B \, \ba \, ,
	\label{eqn:g-rom}
\end{eqnarray}
which can be written componentwise as follows: $\forall \, i = 1 \ldots r,$
\begin{equation}
	\dot{a}_i	
	 = \sum_{m=1}^{r} A_{i m} \, a_m(t)
	 + \sum_{m=1}^{r} \sum_{n=1}^{r} B_{i m n} \, a_n(t) \, a_m(t) \, ,
	\label{eqn:g-rom-2}
\end{equation}
where 
\begin{eqnarray}
	&& A_{im}
	= - Re^{-1} \, \left( \nabla \bphi_m , \nabla \bphi_i \right)\nonumber\\
	&& = - Re^{-1} 
	\sum_{i=1}^{r}  
	\sum_{j=1}^{r} 
	(a_i) ( a_j) 
	\left( \nabla \bphi_{j} ,  \nabla \bphi_{i} \right).\\	
	&& B_{imn}
	= - \bigl( \bphi_m \cdot \nabla \bphi_n , \bphi_i \bigr)\nonumber\\
	&& = - 
	\sum_{i=1}^{r}
	\sum_{m=1}^{r}
	\sum_{n=1}^{r} 
	( a_{m} ) ( a_{n} )(a_i)
	\bigl( \bphi_{m} \cdot \nabla \bphi_{n} , \bphi_{i} \bigr).
	\label{eqn:g-rom-3b}
\end{eqnarray}

\section{Closure Learning}
	\label{sec:closure}

\subsection{Residual Neural Network (ResNet)}
	\label{sec:resnet}

Deep residual neural network (ResNet) has been first introduced for image recognition in \cite{he2016deep}. ResNet has been widely studied and applied in many supervised learning tasks. 
Recent mathematical understanding of deep ResNet has been achieved in the ODE representation of ResNet; for a comprehensive introduction see, e.g., \cite{chang2018reversible,chang2017multi,chen2018neural,lu2017beyond}. 

\begin{figure}[h]
\includegraphics[scale=0.5]{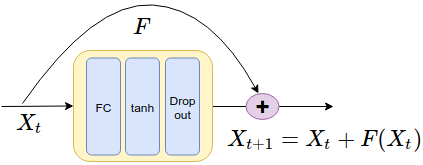}
\caption{ResNet block used in our training.}
\label{fig:resnet}
\end{figure}

To construct the novel ResNet-ROM framework, we consider the ResNet model, which is illustrated in Fig.~\ref{fig:resnet}. 

The input values of forward propagation in the ResNet are given by
\begin{equation}
X_{t+1} = X_{t} + \tanh(W_tX_t+b_t), \ \ t=1,...,N-1,
\label{eqn:resnet}
\end{equation}
where $N$ is the number of layers in the network architecture and $X_t\in\mathcal{R}^{s}$ is the output from each ResNet block. 
The ResNet propagation starts from step $t=1$ with the nonlinear activation function $\tanh$. The initial input layer for the network is $X_0=[Re, t, a_1,...,a_r]^T$. For a standard ResNet in image classification, a convolution layer (CNN) is often included in a residual block. In our work, we use a simplified version of the ResNet which does not include CNN.

\subsection{ROM Closure Modeling}
In realistic nonlinear systems (e.g., convection-dominated flows), current GP-ROMs of the form~\eqref{eqn:g-rom} generally yield inaccurate results, often in the form on numerical instabilities.
This inaccurate behavior is due to the fact that, in order to maintain a low computational cost, GP-ROMs are constructed with a drastically truncated ROM basis $\{ \bphi_1, \ldots, \bphi_r \}$, where often $r = \mathcal{O}(10)$.
In realistic nonlinear systems, this extreme truncation cannot capture the complex nonlinear interactions among the various degrees of freedom of the system.
Thus, current GP-ROMs are computationally efficient, but numerically inaccurate.
To alleviate this inaccurate behavior, GP-ROMs are often supplemented with a ROM closure model~\cite{HLB96,wang2012proper}, i.e., a model  for the important interactions between the ROM basis $\{ \bphi_1, \ldots, \bphi_r \}$ and the discarded ROM modes $\{ \bphi_{r+1}, \ldots, \bphi_{d} \}$, where $d$ is the dimension of the input data.
For example, for the NSE, the standard GP-ROM~\eqref{eqn:g-rom} is generally modified as follows:
\begin{eqnarray}
	\dot{\ba}
	= A \, \ba 
	+ \ba^{\top} \, B \, \ba 
	+ \btau ,
	\label{eqn:g-rom-closure}
\end{eqnarray}
where $\btau$ is a ROM closure model that represents the interactions between the ROM modes and the discarded modes.

We emphasize that the same closure problem needs to be addressed when classical numerical discretization schemes (e.g., finite element or spectral methods) are used in the numerical simulation of turbulent flows.
In those settings, classical discretizations schemes inherently function in the under-resolved regimes (i.e., use coarse meshes or too few spectral modes) and are therefore employing various types of closure models for the unresolved  (e.g., subgrid-scale) information.
These closure models are central to, e.g., large eddy simulation (LES)~\cite{Sag06}, one of the main approaches to the numerical simulation of turbulent flows.
The vast majority of LES closure models have been constructed by using physical insight from Kolmogorov's statistical theory of turbulence.
The concept of energy cascade is central in the development of LES closure models.
The energy cascade states that energy enters the system at the large scales, is transferred to smaller and smaller scales through nonlinear interactions, and is dissipated at the smallest scale (i.e., the Kolmogorov's scale). 
Thus, most LES closure models (e.g., of eddy viscosity type) aim at recovering the energy cascade displayed by the original system (i.e., the NSE).

This physical insight is not available in the ROM setting (see, however,~\cite{CSB03} for a preliminary numerical investigation). 
Thus, current ROM closure models have been deprived of this powerful methodology that represents the core of most LES closure models.

\subsection{ROM Closure Learning}

Our vision is that machine learning represents the perfect means for constructing ROM closure modeling.
Indeed, since physical insight cannot generally be used in a ROM setting, data and machine learning can be utilized instead to develop ROM closure models.
Furthermore, data is used only to construct the ROM closure model; the other ROM operators are built by using the classical Galerkin projection.
Thus, data and machine learning complement (instead of replace) physical based modeling, yielding a hybrid framework that synthesizes the strengths of both approaches.

In this section, we propose a novel ROM closure modeling framework that is constructed by using available data and the ResNet approach described in Section~\ref{sec:resnet}.
The resulting ROM, which we call the {\it residual neural network ROM (ResNet-ROM)}, is schematically illustrated in~\eqref{eqn:cl-rom-schematic} and Fig.~\ref{fig:flow-chart}, and is summarized in Algorithm~\ref{alg:ddf-rom}.

\begin{algorithm}[H]
	\caption{ResNet-ROM}
	\label{alg:ddf-rom}
	\begin{algorithmic}[1]
		\STATE{
					Consider the ROM closure model			
						\begin{eqnarray}
							\dot{\ba} 
							= A \, \ba 
							+ \ba^{\top} \, B \, \ba 
							+ \btau \, .
							\label{eqn:alg-resnet-rom-1}
						\end{eqnarray}
					}
		\STATE{
					Use snapshot data to compute the true vector $\btau$ in~\eqref{eqn:alg-resnet-rom-1}, $\btau^{true}$:
						\begin{eqnarray}
							\tau_i^{true}(t_j)
							&=& - \biggl( \, \overline{\bigl( \bu^{snap}_d(t_j) \cdot \nabla \bigr) \, \bu^{snap}_d(t_j)}^r 
							\nonumber \\
							&& \hspace*{0.3cm} - \bigl( \bu^{snap}_r(t_j) \cdot \nabla \bigr) \, \bu^{snap}_r(t_j)  \, , \, \bphi_i \, \biggr)	,		
							\label{eqn:alg-resnet-rom-2}
						\end{eqnarray}
						where an overbar indicates spatial filtering with ROM projection.
					}
		\STATE{
					Use snapshot data and~\eqref{eqn:alg-resnet-rom-2} to define the approximation function $\btau$ in~\eqref{eqn:alg-resnet-rom-1}, $\btau^{ansatz}$:
						\begin{eqnarray}
							\btau^{ansatz}(t_j)
							= f(\ba, Re, t_{j}) , 
							\label{eqn:alg-resnet-rom-3}
						\end{eqnarray}
						where $f$ is a generic function that needs to be determined.
					}
		\STATE{
					Use ResNet to train the closure term, i.e., to find the form of $f$ in~\eqref{eqn:alg-resnet-rom-3} that is optimal with respect to the minimization problem~\eqref{eqn:loss-function}.
					}

		\STATE{
					The novel ResNet-ROM has the following form:
								\begin{equation}
									\boxed{
									\dot{\ba}
									= A  \, \ba
									+ \ba^{\top} \, B \, \ba
									+ f^{net}(\ba,Re,t)} 
									\label{eqn:alg-resnet-rom-4}
								\end{equation}
								
					}
\end{algorithmic}
\end{algorithm}

In Algorithm~\ref{alg:ddf-rom}, we use the available numerical data and the ResNet approach to construct a ROM closure model, i.e., to find a model for the term $\btau$ in~\eqref{eqn:g-rom-closure}.
Thus, in~\eqref{eqn:resnet}, we make the following choices:
The initial layer contains the Reynolds number, the current time, and the current ROM coefficients in~\eqref{eqn:g-rom-1}, i.e., $X_0 = [Re,t, a_1,...,a_r] \in \mathcal{R}^{(s+2)}$.
The final output layer predicts the values that approximate the closure term, e.g., $X_{N} \approx \btau$.
The optimization problem associated with this network is given by 
\begin{equation}
\min \| \tilde{\btau} - \btau\|^2_F+\lambda R(W, b) ,
\label{eqn:loss-function}
\end{equation}
where the regularizer $R$ penalizes undesirable parameters and can prevent over-fitting, $\tilde{\tau}$ is the output from the neural network, and $W, b$ are weights in the network.

\section{Numerical Experiment}

\subsection{Implementation}

As a test problem for our new approach, we use the 1D Burgers equation, which has been used to test new ROM ideas in simplified settings~\cite{borggaard2011artificial,KV01,xie2018data}:
\begin{eqnarray}
  \left\{\begin{array}{ll}
  u_t-Re^{-1}u_{xx}+{u}u_x=0& x\in\Omega,\\
  u(x,0)=u_0(x)& x\in\Omega,\\
  u(x,t)=0&  x\in\partial\Omega,
  \end{array} \right.
  \label{eqn:burgers}
 \end{eqnarray}
where, for consistency with the notation used for the NSE, the diffusion parameter is denoted as $Re^{-1}$. 
In our numerical tests, $\Omega=[0,1]$ is the computational domain and the time domain is $[0,1]$. 
We use the same initial conditions as those utilized in~\cite{borggaard2011artificial,KV01,xie2018data} $u_0(x)=1,x\in(0,1/2]$, $u_0(x)=0,x\in(1/2,1]$.
These initial conditions yield a steep internal layer that is difficult to capture in the convection-dominated regime that we consider~\cite{borggaard2011artificial,KV01,xie2018data}. 
We first use the  finite element method to generate the FOM data (true solution).
To this end, we utilize a uniform mesh with $N=1024$ grid points and the forward Euler method with $\Delta t=10^{-4}$ for the time discretization.
To construct the ROM basis, we collect $101$ snapshots sampled from $[0,1]$. 
We build the neural network in PyTorch and we train the ROM closure model with a 6 block ResNet with Adams optimizer. 
We perform all the computational experiments on a four core Dell computer with one Geforce GTX GPU and Ubuntu 18.04 system.

\subsection{Reconstruction}
	\label{sec:reconstructive}

\begin{figure}[h]
\includegraphics[scale=0.25]{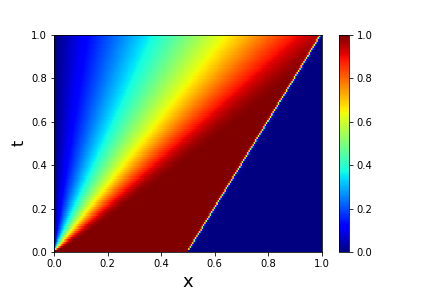}
\includegraphics[scale=0.25]{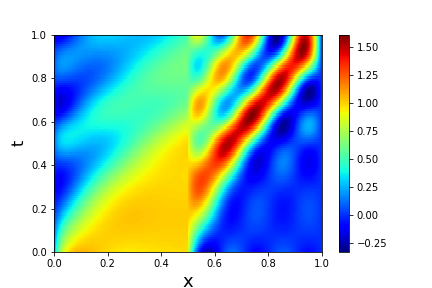}
\includegraphics[scale=0.25]{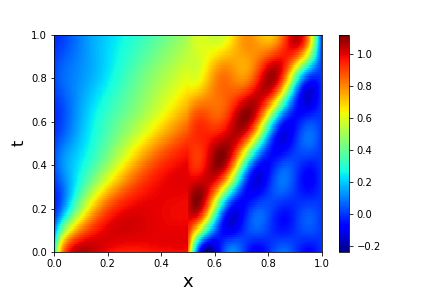}
\hspace*{0.5cm} \includegraphics[scale=0.25]{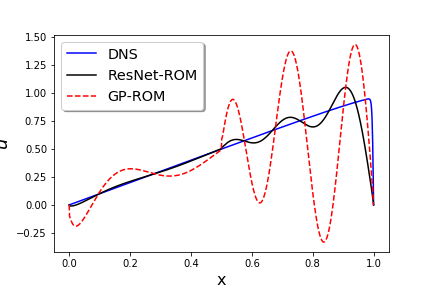}
\caption{
	Reconstructive regime, $Re=1000$:
	FOM (top left), GP-ROM (top right), ResNet-ROM (bottom left), and final time solution for all three simulations (bottom right).
	ResNet-ROM yields the most accurate solution.
	}
\label{fig:recon_re1000}
\end{figure}

In this section, we consider the reconstructive regime, i.e., we  test the ROMs at the same $Re$ as the $Re$ at which the ROMs are constructed.
We choose $Re=1000$ in~\eqref{eqn:burgers} and we use $r=6$ basis functions in all ROMs. 
In Fig.~\ref{fig:recon_re1000}, we plot the solutions of the FOM (top left), GP-ROM (top right), and ResNet-ROM (bottom left). 
When compared with the FOM data, the ResNet-ROM solution is significantly more accurate than the standard GP-ROM solution.
In Fig.~\ref{fig:recon_re1000}, we also plot the FOM, GP-ROM, and ResNet-ROM solutions at the final time step  (i.e., at $t=1$).
This plot shows that the closure term in the ResNet-ROM plays an important role in stabilizing the ROM approximation.
Indeed, the GP-ROM solution displays large, spurious numerical oscillations.
These oscillations are dramatically decreased in the ResNet-ROM solution. 

\begin{figure}[h]
\includegraphics[scale=0.25]{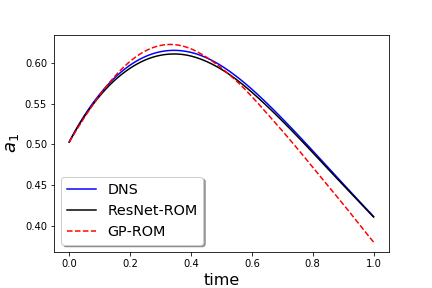}
\includegraphics[scale=0.25]{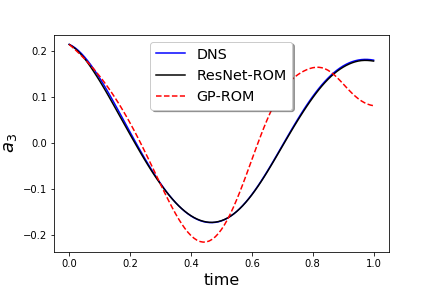}
\caption{
	Reconstructive regime, $Re=1000$:
	Time evolution of ROM coefficients $a_1$ and $a_3$ for FOM, GP-ROM, and ResNet-ROM.
	ResNet-ROM yields the most accurate solution.
	}
\label{fig:recon_timea}
\end{figure}

In Fig.~\ref{fig:recon_timea}, we plot the the time evolution of the ROM coefficients $a$ for the FOM, GP-ROM, and ResNet-ROM.
The plots show that the ResNet-ROM is significantly more accurate than the standard GP-ROM.


\subsection{Prediction}
	\label{sec:predictive}
	
To study the robustness of the new ResNet-ROM, we test its predictability, i.e., we train the ResNet-ROM closure term on data from multiple $Re$ and we then test the ResNet-ROM to predict the ROM dynamics at different $Re$. 
The training data space is sampled at $Re=[20, 50, 100, 200, 500, 800, 1000]$, and the test data contains $Re=[30, 80, 300, 1200]$. 
Note that the solution of Burgers equation is affected by $Re$. 
Small $Re$ values yield a slow movement of the sharp internal layer, while large $Re$ values can speed up this movement. 

\begin{figure*}[h]
\includegraphics[scale=0.25]{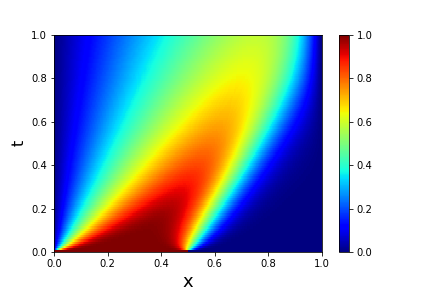}
\includegraphics[scale=0.25]{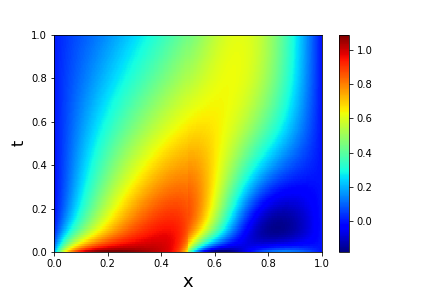}
\includegraphics[scale=0.25]{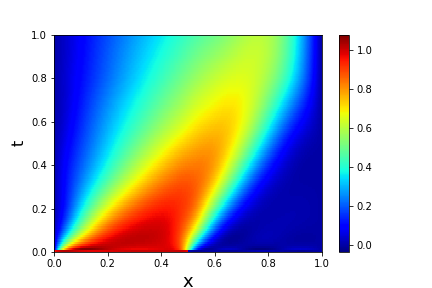}
\includegraphics[scale=0.25]{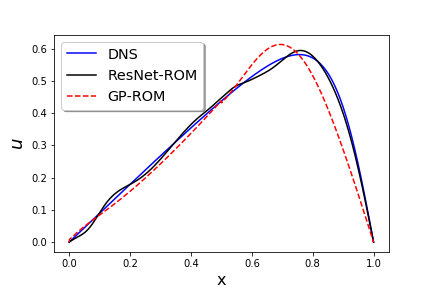}\\
\includegraphics[scale=0.25]{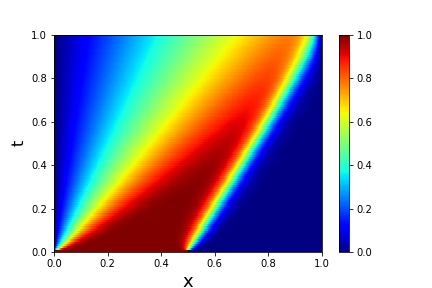}
\includegraphics[scale=0.25]{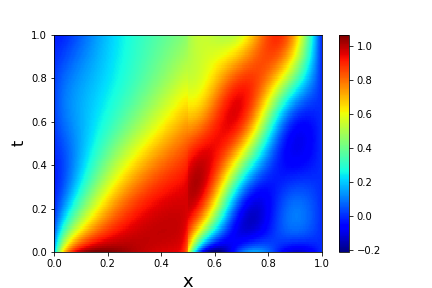}
\includegraphics[scale=0.25]{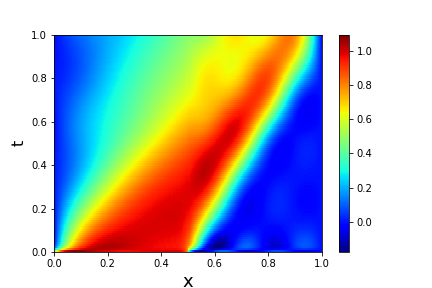}
\includegraphics[scale=0.25]{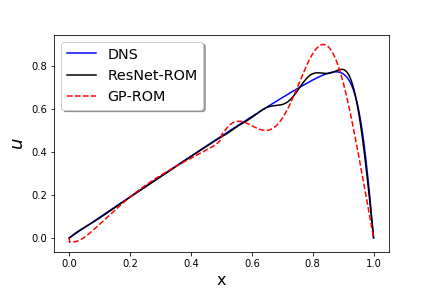}\\
\includegraphics[scale=0.25]{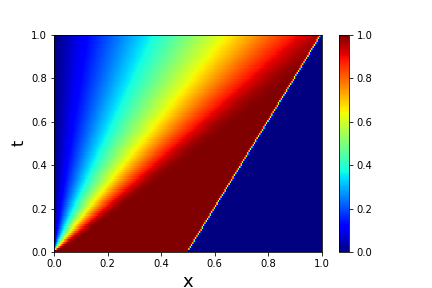}
\includegraphics[scale=0.25]{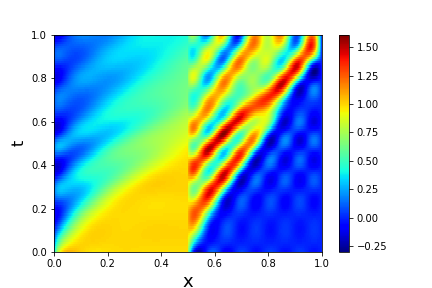}
\includegraphics[scale=0.25]{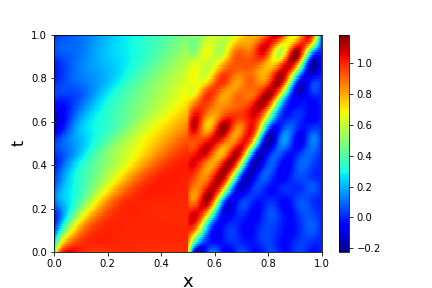}
\includegraphics[scale=0.25]{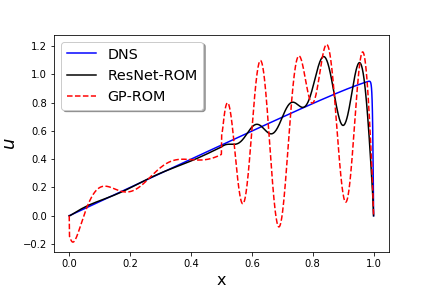}
\caption{
	Predictive regime:
	ROMs are trained on data from $Re=[20, 50, 100, 200, 500, 800, 1000]$ and are tested at $Re=30$ (first row), $Re=80$ (second row), $Re=1200$ (third row). 
	Results presented for FOM (first column), GP-ROM (second column), ResNet-ROM (third column), and final time solution for all three simulations (fourth column).
	}
\label{fig:pred_mulre}
\end{figure*}

In Fig.~\ref{fig:pred_mulre}, for $Re = 30, 80$, and $1200$ (which are different from the training $Re$ values), we plot the solutions for the FOM (first column), GP-ROM (second column), ResNet-ROM (third column), and final time solution for all three simulations (fourth column). 
These plots show that the ResNet-ROM is consistently the most accurate ROM, especially for the largest $Re$  value.
In Fig.~\ref{fig:pred_mulre}, we also plot the FOM, GP-ROM, and ResNet-ROM solutions at the final time step  (i.e., at $t=1$).
These plots show that the closure term in the ResNet-ROM plays an important role in stabilizing the ROM approximation.

\begin{figure}[h]
\includegraphics[scale=0.25]{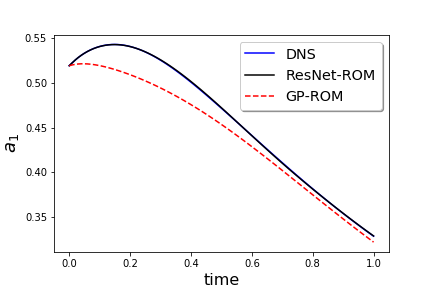}
\includegraphics[scale=0.25]{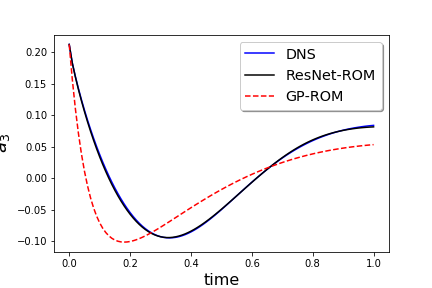}\\
\includegraphics[scale=0.25]{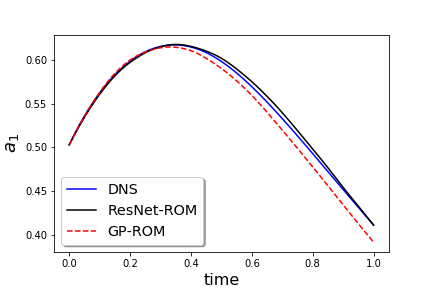}
\includegraphics[scale=0.25]{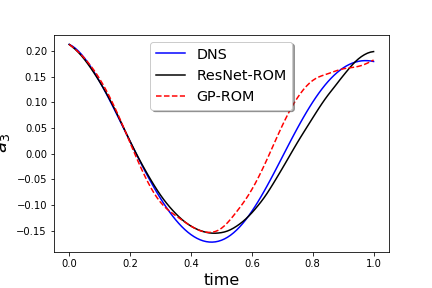}
\caption{
	Predictive regime:
	Time evolution of ROM coefficients $a_1$ and $a_3$ for FOM, GP-ROM, and ResNet-ROM. Results for $Re=30$ (top), $Re=1200$ (bottom).
	ResNet-ROM yields the most accurate solution. 
	}
\label{fig:time_evolve_pred}
\end{figure}

In Fig.~\ref{fig:time_evolve_pred}, we plot the the time evolution of the ROM coefficients $a$ for the FOM, GP-ROM, and ResNet-ROM.
These plots show that the ResNet-ROM is significantly more accurate than the standard GP-ROM.

Overall, we draw the same conclusion as in the reconstructive regime (Section~\ref{sec:reconstructive}):
In all cases, the ResNet-ROM is significantly more accurate than the standard GP-ROM.

\subsection{Comparison}

In this section, in the numerical simulation of the Burgers equation, we also make a numerical comparison between the new ResNet-ROM and several modern ROM closure models:
the POD artificial viscosity model (POD-AV)~\cite{borggaard2011artificial}, the POD Smagorinsky model (POD-L)~\cite{akhtar2012new}, the evolve-then-filter ROM (EF-ROM)~\cite{wells2017evolve}, the approximate deconvolution ROM (AD-ROM)~\cite{xie2017approximate}, and the data-driven filtered  ROM (DDF-ROM)~\cite{xie2018data}.
In Table~\ref{table:compare}, we list the $L^2$ errors of the ResNet-ROM and and the other closure models.
These results show that the ResNet-ROM is at least an order of magnitude lower than the errors of the other closure models.

\begin{table}[h]
\caption{$L^2$ errors of the new ResNet-ROM and other closure models.} \label{table:compare}
\begin{center}
\begin{tabular}{ll}
\textbf{Model}  &\textbf{$L^2$ Error} \\
\hline \\
\textbf{CL-ROM}                             &\textbf{0.0004324}\\
POD-AV (\cite{borggaard2011artificial})         &0.0090\\
POD-L (\cite{akhtar2012new})             &0.0173 \\
EF-ROM(\cite{wells2017evolve})            &0.0699 \\
AD-ROM(\cite{xie2017approximate})            &0.0633 \\
DDF-ROM(\cite{xie2018data})            &0.0627 \\
\end{tabular}
\end{center}
\end{table}

\section{Conclusions}

In this paper, we used available data and deep residual neural networks (ResNet) to construct a novel reduced order model (ROM) closure for complex nonlinear settings.
We emphasize that the ResNet-ROM closure terms are much more general than the ansatzes generally used in ROM closure modeling~\cite{xie2018data}.
We tested the novel ResNet-ROM in the numerical simulation of the Burgers equation.
For comparison purposes,  we investigated the standard Galerkin projection ROM (GP-ROM) and the full order model (FOM).
We considered two settings:
(i) a reconstructive regime, in which the Reynolds number $Re$ is the same in the training and testing stages;  and 
(ii) a predictive regime, in which $Re$ used in the testing regime is different from the $Re$ used in the training regime.
In both regimes, the new ResNet-ROM was consistently more accurate than the standard GP-ROM.
Furthermore, the ResNet-ROM was also dramatically  more accurate than several other ROM closure models from the literature.

There are several research directions that we plan to pursue:
We will test the novel ResNet-ROM on more challenging test problems (e.g., 3D turbulent flows) and we will  compare it with state of the art closure models.
We will also investigate alternative approaches to develop the ROM closure term $\btau$.
Indeed, in this paper we used the  ROM projection as a spatial filter in the construction of the ROM closure term $\btau$.
We plan to investigate different ROM spatial filters, such as the differential filter~\cite{wells2017evolve}.
This different ROM filter will yield a different ROM closure term $\btau$ and, therefore, a different ResNet-ROM.

\bibliographystyle{plain}
\bibliography{xie}
\end{document}